\documentclass[pdflatex]{sn-jnl}

\usepackage{graphicx}%
\usepackage{multirow}%
\usepackage{makecell}

\usepackage{amsmath,amssymb,amsfonts}%
\usepackage{amsthm}%
\usepackage{mathrsfs}%
\usepackage[title]{appendix}%
\usepackage{xcolor}%
\usepackage{textcomp}%
\usepackage{manyfoot}%
\usepackage{booktabs}%
\usepackage{algorithm}%
\usepackage{algorithmicx}%
\usepackage{algpseudocode}%
\usepackage{listings}%
\usepackage{amssymb} 

\theoremstyle{thmstyleone}%
\theoremstyle{thmstyletwo}%

\theoremstyle{thmstylethree}%

\usepackage{lineno}

\begin{document}

\title[LoRA-based methods on Unet for transfer learning in aneurysmal subarachnoid hematoma segmentation]{LoRA-based methods on Unet for transfer learning in aneurysmal subarachnoid hematoma segmentation}


\author*[1]{\fnm{Cristian} \sur{Minoccheri}}\email{minoc@umich.edu}
\equalcont{These authors contributed equally to this work.}

\author*[1]{\fnm{Matthew} \sur{Hodgman}}\email{hodgman@umich.edu}
\equalcont{These authors contributed equally to this work.}

\author[1]{\fnm{Haoyuan} \sur{Ma}}
\email{mhaoyuan@umich.edu}

\author[2]{\fnm{Rameez} \sur{Merchant}}
\email{rmrdq@health.missouri.edu}

\author[1]{\fnm{Emily} \sur{Wittrup}}\email{ewittrup@umich.edu}

\author[2]{\fnm{Craig} \sur{Williamson}}
\email{craigaw@med.umich.edu}

\author[1,3,4]{\fnm{Kayvan} \sur{Najarian}}\email{kayvan@umich.edu}

\affil[1]{\orgdiv{Gilbert S. Omenn Department of Computational Medicine and Bioinformatics},  \orgname{University of Michigan}, \orgaddress{\street{1109 Geddes Avenue}, \city{Ann Arbor}, \postcode{48104}, \state{MI}, \country{USA}}}

\affil[2]{\orgdiv{Department of Neurosurgery}, \orgname{University of Michigan}, \orgaddress{\street{1109 Geddes Avenue}, \city{Ann Arbor}, \postcode{48104}, \state{MI}, \country{USA}}}

\affil[3]{\orgdiv{Department of Emergency Medicine}, \orgname{University of Michigan}, \orgaddress{\street{1109 Geddes Avenue}, \city{Ann Arbor}, \postcode{48104}, \state{MI}, \country{USA}}}

\affil[4]{\orgdiv{Department of Electrical Engineering and Computer Science}, \orgname{University of Michigan}, \orgaddress{\street{1109 Geddes Avenue}, \city{Ann Arbor}, \postcode{48104}, \state{MI}, \country{USA}}}


\abstract{\textbf{Background:}
Aneurysmal subarachnoid hemorrhage (SAH) is a life-threatening neurological emergency with mortality rates exceeding 30\%. While deep learning techniques show promise for automated SAH segmentation, their clinical application is limited by the scarcity of labeled data and challenges in cross-institutional generalization. Transfer learning from related hematoma types represents a potentially valuable but underexplored approach. Although Unet architectures remain the gold standard for medical image segmentation due to their effectiveness on limited datasets, Low-Rank Adaptation (LoRA) methods for parameter-efficient transfer learning have been rarely applied to convolutional neural networks in medical imaging contexts. The importance of SAH diagnosis and the time-intensive nature of manual annotation would benefit from automated solutions that can leverage existing multi-institutional datasets from more common conditions.

\textbf{Methods:} We implemented a Unet architecture pre-trained on computed tomography scans from 124 traumatic brain injury patients across multiple institutions, then fine-tuned on 30 aneurysmal SAH patients from the University of Michigan Health System using 3-fold cross-validation. We developed a novel CP-LoRA method based on tensor canonical polyadic (CP) decomposition and introduced DoRA variants (DoRA-C, convDoRA, CP-DoRA) that decompose weight matrices into magnitude and directional components. We compared these approaches against existing LoRA methods (LoRA-C, convLoRA) and standard fine-tuning strategies across different modules on a multi-view Unet model. Performance was evaluated using Dice scores stratified by hemorrhage volume, with additional assessment of predicted versus annotated blood volumes.

\textbf{Results:}  Transfer learning from traumatic brain injury to aneurysmal SAH demonstrated feasibility with all fine-tuning approaches achieving superior performance compared to no fine-tuning (mean Dice 0.410±0.26). The best-performing traditional approach was decoding module fine-tuning (Dice 0.527±0.20). LoRA-based methods consistently outperformed standard Unet fine-tuning, with DoRA-C at rank 64 achieving the highest overall performance (Dice 0.572±0.17). Performance varied by hemorrhage volume, with all methods showing improved accuracy for larger volumes (Dice 0.682-0.694 for volumes $>$100mL vs. Dice 0.107-0.361 for volumes $<$25mL). CP-LoRA achieved comparable performance to existing methods while using significantly fewer parameters. Over-parameterization with higher ranks (64-96) consistently yielded better performance than strictly low-rank adaptations.

\textbf{Conclusions:} This study demonstrates that transfer learning between hematoma types is feasible and that LoRA-based methods significantly outperform conventional Unet fine-tuning for aneurysmal SAH segmentation. The novel CP-LoRA method offers parameter efficiency advantages, while DoRA variants provide superior segmentation accuracy, particularly for small-volume hemorrhages. The finding that over-parameterization improves performance challenges traditional low-rank assumptions and suggests clinical applications may benefit from higher-rank adaptations. These results support the potential for automated SAH segmentation systems that leverage large multi-institutional traumatic brain injury datasets, potentially improving diagnostic speed and consistency when specialist expertise is unavailable.
}




\maketitle

\section{Background}\label{sec1}

A subarachnoid hematoma (SAH) is clotted blood on the surface, ventricles and basal cisterns of the brain in the subarachnoid space, between the pia and dura mater. Spontaneous, non-traumatic subarachnoid hemorrhage is most often the result of aneurysm rupture \cite{Van_Gijn2007-zb}. SAH cases are responsible for as much as 10\% of all strokes \cite{Rincon2013-jy} and exhibit mortality rates over 30\% \cite{Nieuwkamp2009-ma}. On average, patients with SAH are generally younger than those with other types of intracranial hemorrhages and strokes (59 years compared to 73 and 81, respectively) \cite{Johnston1998-wn}. The severe burden caused by SAHs, the importance of quick intervention, and the current time-consuming manual annotation process, has prompted advancements in automated, artificial intelligence-based methods. Many studies have demonstrated the efficacy of deep learning for segmenting various types of hematomas \cite{Yao2020-fi, Yu2022-qa, Farzaneh2017-eu, Ironside2019-kw}. Automated hematoma segmentation enables (1) fast and accurate hematoma detection (2) standardized quality of care when specialists are unavailable (3) feature extraction for quantifying risk of complications. However, few studies focus on aneurysmal SAH segmentation, which has proven to be a more difficult task due to many small pockets of blood along the brain's surface \cite{Street2023-ft, Barros2020-nm, Boers2014-mi, Garcia2023-uo, Butler2025-yt, Kiewitz2024-mk, Sanchez2025-oi}.

Previous efforts to develop automatic aneurysmal SAH segmentation models exhibit varying performance. We report the computational methodology, cohort size, aggregate Dice score, and whether the study used data from multiple institutions in Table \ref{tab:studies}. These studies leverage a variety of methods to segment--mostly aneurysmal--SAH on datasets of very different sizes, with highly variable performance. All but one of these studies exclusively use aneurysmal SAH cases from a single institution for model training. \cite{Yuce2025-kf} used a multi-institutional dataset but report poor performance (mean Dice=0.35) While unexplored to date, using multi-institutional data in addition to transfer learning may improve both generalizability and performance.

\begin{table}[t]
\centering
\caption{Overview of previous studies on automatic SAH segmentation.}
\label{tab:studies}
\begin{tabular}{
    >{\raggedright\arraybackslash}p{2.3cm} 
    >{\raggedright\arraybackslash}p{3.5cm} 
    >{\centering\arraybackslash}p{0.8cm} 
    >{\centering\arraybackslash}p{0.7cm} 
    >{\centering\arraybackslash}p{1.7cm}
    >{\centering\arraybackslash}p{1.4cm}
}
\toprule
\textbf{Study} & \textbf{Method} & \textbf{Cohort Size} & \textbf{Dice} & \textbf{Multi-Institution} & \textbf{Transfer Learning}\\
\midrule
Boers et al. \cite{Boers2014-mi} & Atlas-based approach & 30 & 0.55 & & \\
Barros et al. \cite{Barros2020-nm} & CNN & 268 & 0.63 & & \\
Street et al. \cite{Street2023-ft} & ITK-SNAP, Random Forest & 42 & 0.92 & & \\
Garcia et al. \cite{Garcia2023-uo} & Transformer-based Unet & 80 & 0.74 & & \\
Butler et al. \cite{Butler2025-yt} & Rule-based segmentation & 20 & 0.43 & & \\
Kiewitz et al. \cite{Kiewitz2024-mk} & 2D nnU-Net & 73 & 0.616 & & \\
Lin et al. \cite{Lin2025-yx} & ResUNet with attention & 1347 & 0.93 & & \\
Yüce et al. \cite{Yuce2025-kf} & 3D U-Net & 353 & 0.35 & \checkmark & \\
Sanchez et al. \cite{Sanchez2025-oi} & K-means clustering & 500 & 0.602 & & \\
\bottomrule
\end{tabular}
\end{table}

Transfer learning is a common, deep-learning model development approach that uses knowledge accumulated from one problem to solve a different, related problem. Often in practice, models are trained to accomplish a general task on a large dataset, then "fine-tuned" on a smaller, more specialized task and dataset. Transfer learning is widely employed in medical image processing because large, labelled datasets are scarce and it can greatly improve model performance \cite{Van_Opbroek2015-zh}. Tuning models originally trained on medical images, instead of out-of-domain images like ImageNet, shows the best performance \cite{Atasever2023-gb}. Following this training strategy, using imaging data from other, more common hematomas, like those caused by traumatic brain injury (TBI) may improve aneurysmal SAH segmentation performance. Various fine-tuning strategies for Unet architectures exist, focusing on which part of the Unet model to freeze, and the optimal solution is an open question \cite{Amiri2020}.

LoRA (Low-Rank Adaptation) \cite{lora} is a parameter-efficient fine-tuning method that freezes the original model weights and injects trainable low-rank matrices into the network to approximate weight updates. More recently, the variant DoRA (Weight-Decomposed Low-Rank Adaptation) \cite{dora} has been proposed: DoRA decomposes the full weight as a magnitude vector and a directional matrix (independently updated), enabling more expressive updates and improving stability. LoRA and DoRA have been widely employed in transformer architectures allows to greatly reduce the number of trainable parameters while maintaining - and often improving - performance; however, applications of LoRA to CNNs are limited to convLoRA \cite{convlora} and LoRA-C \cite{lorac}, and applications of DoRA are completely absent.

There are several practical and architectural reasons why Unet models are often preferred over transformers for hematoma segmentation and similar medical imaging tasks: (1) Unet performs well on small datasets, which is common in medical imaging. (2) It is built on convolutions, which naturally encode translation equivariance and local spatial priors.  (3) It is lightweight and computationally efficient. (4) Unet’s encoder–decoder structure with skip connections is well-suited for dense pixel-wise segmentation, preserving both local detail and global context. Therefore, Unet remains a strong choice for medical image segmentation, but LoRA methods for transfer learning this architecture have not been adequately explored.

To assess the efficacy of transfer learning to develop an aneurysmal SAH segmentation model, we trained a standard Unet \cite{Unet} and a multi-view Unet \cite{Yao2020-fi} on brain computed tomography (CT) scans of 124 patients with TBI from multiple institutions, followed by fine-tuning on CT scans from aneurysmal SAH patients from the University of Michigan Health System. Given a limited dataset of 30 SAH patients, we perform 3-fold cross-validation for fine-tuning and testing. We introduce a novel LoRA technique, CP-LoRA, and add the DoRA framework to LoRA techniques, introducing DoRA-C, convDoRA, and CP-DoRA. We compare these LoRA-based methods on the classic Unet model and show that they outperform standard fine-tuning strategies for the multi-view Unet architecture for our dataset. We use the multi-view Unet as a non-trivial benchmark for the proposed LoRa/DoRA techniques. Standard fine-tuning on the classic Unet consistently underperformed that on the multi-view Unet; therefore it is excluded from our analysis. On the other hand, by applying LoRA/DoRA methods to a baseline Unet we intend to underline how similar results should also translate to the many variants of Unet, including 3D Unet. Our results demonstrate the potential of transfer learning between hematoma types using LoRA methods. The main contributions of the paper are as follows:

1) Test the feasibility of transfer learning across hematoma types with a limited amount of training data, in the particularly challenging case of aneurysmal SAH;

2) Introduce a novel low rank method for Unet, CP-LoRA, that uses fewer parameters than existing methods LoRA-C and convLoRA;

3) Introduce DoRA variants of LoRA methods: DoRA-C, convDoRA, CP-DoRA;

4) Provide a comparison of all LoRA/DoRA methods on a Unet model, show that they outperform fine-tuning in the case of a multi-view Unet with non-trivial skip connections, and show that over-parametrization provides improved performance compared to low rank adaptations.

\section{Methods}\label{sec2}

\subsection{Multi-view Unet}

\begin{figure}[t]
    \centering
    \includegraphics[width=1.0\columnwidth]{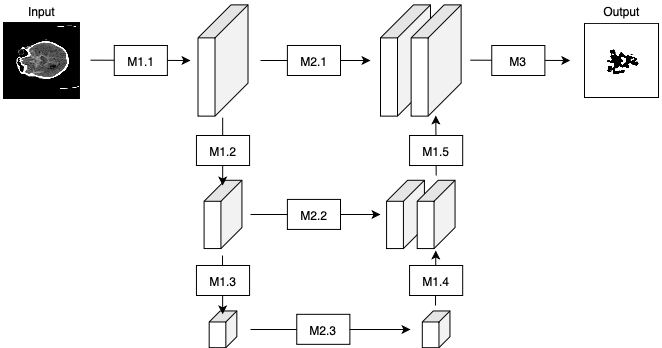}
    \caption{Two-level multi-view Unet architecture.}
    \label{fig:multiview}
\end{figure}

In this work, we compare LoRA-based methods on the original Unet architecture \cite{Unet} with a multi-view Unet developed by Yao et al. \cite{Yao2020-fi}, which showed good performance at hematoma segmentation for patients with TBI. Unet models have shown state of the art performance on medical image segmentation \cite{Unet-review}, and many variants of the original Unet architecture have been proposed. The multi-view model is a variation of the standard Unet where skip-connections include dilated convolutional layers to extract features at multiple scales, resulting in more fine-grained feature maps, and the loss function is weighted average of metrics over images from different contrast enhancements to improve the generalization of the network. For a full description of the model, see the original publication \cite{Yao2020-fi}. In this study, we used a 2-level version of the model, implemented in PyTorch. The multi-view Unet and standard Unet models used in this work can be structurally compared in Figures \ref{fig:multiview} and \ref{fig:unet}, respectively.

Figure \ref{fig:multiview} represents a schematic diagram of the multi-view Unet. There are three types of convolutional blocks: M1, M2, and M3. In the diagram, different blocks of type M1 (respectively M2) are denoted as M1.i (respectively M2.i). M1 consists of two convolutional layers with 3x3 filters and ReLU activation functions. M2 consists of three consecutive dilated convolutional layers, with 3x3 filters and dilation rates of 1, 2, and 4, and a skip connection concatenating the input and the output. M3 consists of two convolutional layers with filters of size of 3x3, and one layer with a 1x1 filter. The mask is generated as a probability map with a pixel-wise softmax activation function.

The basic loss function for an image is given by the following formula, where $\Omega$ is the image pixel grid, $x \in \Omega$ is a pixel location, $L(x)$ is the annotated mask for pixel $x$, and $O_0(x)$ is the output probability of the network for pixel $x$: 

\begin{equation}
    \text{loss}=-\sum_{x \in \Omega}\frac{L(x)O_0(x)}{L(x)+O_0(x)}.
    \label{eq:loss}
\end{equation}
\vspace{0pt}

However, to make the model more robust to CT scans obtained under different conditions from multiple centers with different acquisition and imaging protocols, multiple images $I_0, I_1, \ldots, I_N$ are generated with different contrast settings, and a mixed loss function is computed by adding the loss of each of these images. The losses are scaled by weights $w_0, w_1, \ldots, w_N$ so that images very different from $I_0$ have lower weight, proportionally to the amount of noise added: 

\begin{equation}
    \text{mixed loss}=-\sum_{i=0}^N w_i\sum_{x \in \Omega}\frac{L(x)O_i(x)}{L(x)+O_i(x)}.
    \label{eq:mixedloss}
\end{equation}

\subsection{LoRA and DoRA methods for CNNs}
LoRA \cite{lora} decomposes weight updates in linear layers as:

\begin{equation}
    W = W_0 + \frac{\alpha}{R} BA
    \label{eq:lora}
\end{equation}
\vspace{0pt}

where $W_0$ is the pretrained frozen layer, $B \in \mathbb{R}^{d \times r}$, $A \in \mathbb{R}^{r \times k}$, $r \ll \min(d,k)$, and $\alpha$ is a scaling LoRA parameter. Its variant DoRA \cite{dora} decomposes the full weight matrix as a magnitude vector $m=\|W_0\|_c$ (where $\| \cdot \|_c$ column-wise norms) and a directional matrix $\frac{W_0}{\|W_0\|_c}$; $m$ is trainable and directly updated, while LoRA updates are applied to the directional matrix. Therefore, DORA can be formulated as 

\begin{equation}
    \underline{m}\cdot \frac{W_0+\underline{B}\underline{A}}{\|W_0+\underline{B}\underline{A}\|_c}
    \label{eq:dora}
\end{equation}
\vspace{0pt}

where the underlined variables are trainable.

In CNNs, weights typically have 4D shape ${c_{out} \times c_{in} \times  k \times r}$, where $c_{out}$ is the number of output channels, $c_{in}$ is the number of input channels, and $k$ is the kernel size. There are therefore several ways to define tensors $A$ and $B$ and to multiply them together. 
LoRA-C \cite{lorac} defines $A \in \mathbb{R}^{R \times c_{in} \times k}$, $B \in \mathbb{R}^{c_{out} \times k \times R}$;
the resulting update is $W = W_0 + B \star A$, where $R$ is the chosen rank hyperparameter, and $\star$ denotes the tensor contraction along the rank dimension $R$.  The authors report best performance for $\alpha = 2R$; this is a reasonable value for any LoRA method, and we will adopt it as well.  The number of parameters for each layer is $(c_{in}+c_{out})kR$.

ConvLoRA \cite{convlora} defines $A \in \mathbb{R}^{R \times c_{in} \times k_h \times k_w}$ and $B \in \mathbb{R}^{c_{out} \times R}$; the resulting update is similarly 

\begin{equation}
    W = W_0 + \frac{\alpha}{R} (B \star A).
    \label{eq:convlora}
\end{equation}
\vspace{0pt}

The number of parameters for each layer is ${(c_{in}k^2+c_{out})R}$. The authors suggest using $R=2$ for $3 \times 3$ convolutions, and do not add bias terms.

We introduce a third LoRA version, CP-LoRA, based on the tensor CP-decomposition (see for eample \cite{tensor-survey} for an overview), that introduces fewer parameters than both convLoRA and LoRA-C (for the same rank $R$). This version automatically extends to higher order arrays (which could prove useful in low parameter fine-tuning of 3D CNNs); therefore we define it in general.
A tensor $X$ of shape $d_1 \times \ldots \times d_N$ is an array of order $N$ (i.e., with $N$ modes), and its entry in position $(i_1,\ldots,i_N)$ is denoted by $X_{i_1,\ldots,i_N}$. A rank $1$ tensor of shape $d_1 \times \ldots \times d_N$ is the outer product (denoted $\circ$) of $N$ vectors $a_1 \circ \ldots \circ a_N$, with $a_i \in \mathbb{R}^{d_i}$ and $(a_1 \circ \ldots \circ a_N)_{i_1,\ldots,i_N}=(a_1)_{i_1} \cdot \ldots \cdot (a_N)_{i_N}$. A CP decomposition of rank $r$ approximates a tensor with a sum of $R$ rank $1$ terms 

\begin{equation}
    X=\sum_{i=1}^R a_i^{(1)} \circ \ldots \circ a_i^{(N)}.
    \label{eq:cp}
\end{equation}
\vspace{0pt}

This is an higher order version of the low rank approximation of matrices. Therefore, it is natural to consider LoRA updates of a frozen weight tensor $W_0$ of shape $d_1 \times \ldots \times d_N$ in the form of 

\begin{equation}
    W_0+\sum_{i=1}^R a_i^{(1)} \circ \ldots \circ a_i^{(N)}.
    \label{eq:loraupdate}
\end{equation}
\vspace{0pt}

The number of parameters for each layer is ${(c_{in}+c_{out}+2k)R}$. 

We additionally introduce DoRA versions of the previous LoRA methods: DoRA-C, convDoRA, CP-DoRA. In each case, following the DoRA intuition, we create a vector $m$ of length $c_{out}$ obtained by normalizing each slice for a fixed output channel. We then apply LoRA-C, convLoRA, or CP-LoRA to the resulting normalized weight tensor as before; in the case of DoRA-C and convDoRA, for example, we obtain 

\begin{equation}
    \underline{m}\cdot \frac{W_0+\underline{B}\star \underline{A}}{\|W_0+\underline{B}\star \underline{A}\|_{c_{out}}}.
    \label{eq:doraupdate}
\end{equation}
\vspace{0pt}

LoRA methods are typically used with small values of $R$ to leverage the low-rank structure in weight updates during fine-tuning — allowing large pretrained models to adapt efficiently to new tasks without modifying most of their original parameters. However, these methods also allow for over-parametrization by using a large rank, injecting even more parameters than with full fine-tuning, which allows the model to be more expressive. While a higher rank can be inefficient and increase the risk of overfitting, it has been shown empirically how it can sometimes lead to improved performance and reduced variation.

\begin{figure}[t]
    \centering
    \includegraphics[width=1.0\columnwidth]{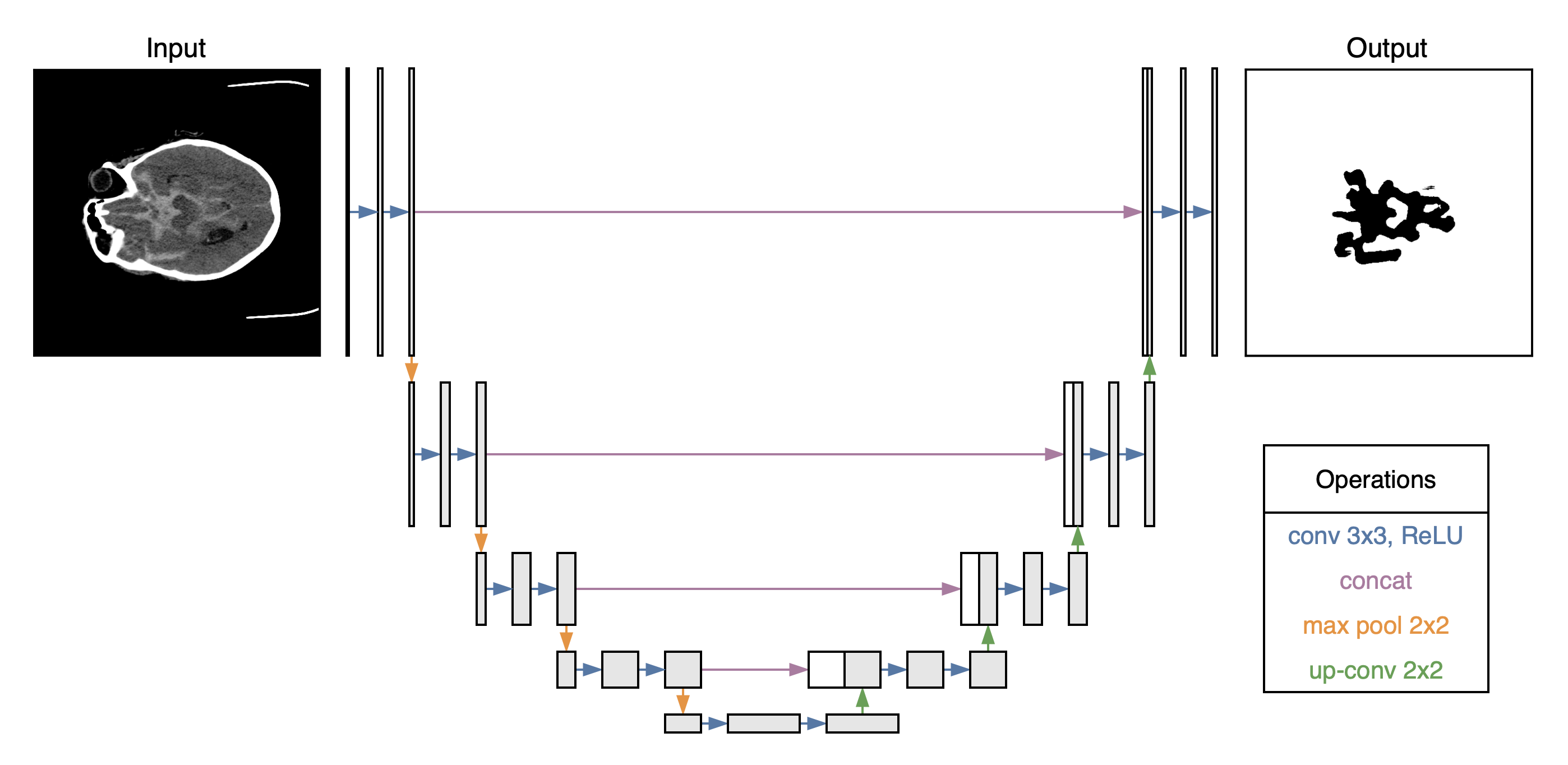}
    \caption{Standard Unet architecture.}
    \label{fig:unet}
\end{figure}

In this work, we apply LoRA/DoRA methods to standard Unet because it is a classic architecture with many variants and represents a standard in image segmentation. We compare the proposed methods against the multi-view Unet because it represents a non-trivial benchmark of performance. For each convolution operation in the Unet model (see Figure \ref{fig:unet}), we define LoRA/DoRA weight matrices that, when combined together, form a weight tensor of the same dimensions as the convolution layer weight tensor. The LoRA/DoRA weights are ignored during pretraining and then fit and added to the original layer weights during tuning. By showing results on the baseline we intend to underline how similar results should also translate to the many variants of Unet, including 3D Unet.

\subsection{Data Preprocessing and Model Development}
We pre-train the Unet models using the same technique as described in the original publication \cite{Yao2020-fi}, on axial brain CT scans from 112 patients and validated its performance on 12 patients. This cohort includes 71 patients with TBI from the multi-center PROTECT III trial \cite{Wright2014-qq}, 28 patients with traumatic brain injury from the University of Michigan Health System, and 25 patients without any brain injury from the University of Michigan Health System. These scans were hand-annotated by a radiologist.

For model tuning and evaluation, we used 30 day-1 axial CT brain scans of patients with aneurysmal SAH admitted to the University of Michigan from 2016 to 2019. A radiologist hand-annotated the SAH boundaries on each scan. For more details on the University of Michigan Subarachnoid Hemorrhage Database used for this study, see  \cite{Nguyen2021-pk}.

All CT scans underwent the same preprocessing as outlined in detail in \cite{Yao2020-fi}, including conversion to Hounsfield units, re-orientation so the brain is horizontal, cropping/padding to 512x512 pixel resolution, and removal of neck CT slices.

We augmented the training and tuning data with contrast adjustment, random horizontal flipping, and elastic transformation as described in \cite{Yao2020-fi}. This resulted in the data falling into four near equally-sized types: not augmented, horizontally flipped, elastic transformed, and both horizontally flipped and elastic transformed. Testing data underwent only contrast adjustment before evaluation.

To tune and evaluate the models, we employ a 3-fold cross validation approach using the 30 CT scans dataset. SAH scans are randomly divided in three non-overlapping sets and we iteratively use two sets to fine-tune an instance of the model and one held-out set for testing. This results in three distinct model instances and their performance on three subsets of the data, enabling denoised, aggregate performance evaluation.

\section{Results}

We pre-trained the multi-view Unet and the standard Unet on TBI hematoma CT scans from mulitple institutions and tuned them on SAH CT scans from the University of Michigan via 3-fold cross validation.

The 2-level multi-view Unet and the standard Unet were trained using the TBI dataset for 60 epochs, with a learning rate of 0.001, and a batch size of 2. We used the same hyperparameters for fine-tuning, save tuning for only 20 epochs to avoid overfitting.

Because the aneurysmal SAH dataset only includes 30 patients, we performed 3-fold cross validation for tuning and evaluation. We pre-trained a Unet model and fine-tuned the entire model as a state-of-the-art comparison \cite{Amiri2020, Unet}. To assess the performance of the different tuned models, we used the Dice score 


\begin{equation}
    \text{Dice score} = \frac{2\text{TP}}{2\text{TP}+\text{FN}+\text{FP}}
    \label{eq:dice}
\end{equation}
\vspace{0pt}

where TP is the number of correctly segmented pixels, FN is the number of SAH pixels not segmented, and FP is the number of segmented background pixels, according to the clinician-labeled ground truth. We computed the annotated blood volume as ${n \times \textit{slice thickness} \times \textit{pixel width} \times \textit{pixel height}}$ as reported in the CT image metadata, with $n$ being the number of pixels of annotated hematoma. The tensor operations are used in the Python implementation Tensorly \cite{Kossaifi2019}.

\begin{table}[t]\centering
\caption{The modules of the multi-view Unet to be fine-tuned for each strategy.}\label{tab:modules}
\begin{tabular}{lrr}\toprule
Fine-tuning Strategy &Modules \\\midrule
Shallow &M1.1, M3 \\
Deep &M1.3, M1.4 \\
Encoding &M1.1, M1.2, M1.3 \\
Decoding &M1.4, M1.5, M3 \\
All & M1.*, M2.*, M3 \\
\bottomrule
\end{tabular}
\end{table}

\begin{table}[t]\centering
\caption{Number of trainable parameters for each fine-tuning method using rank $r$.}\label{tab:params}
\begin{tabular}{lr}\toprule
Method & Parameters \\\midrule
CP-LoRA & $r \cdot 21,633$ \\
CP-DoRA & $r \cdot 29,441$ \\
convLoRA & $r \cdot 64,138$ \\
convDoRA & $r \cdot 71,946$ \\
DoRA-C & $r \cdot 55,621$ \\
LoRA-C & $r \cdot 47,813$ \\
\bottomrule
\end{tabular}
\end{table}


\begin{table*}[t]\centering
\caption{Average patient Dice scores for the multi-view Unet model and different fine-tuning strategies on aneurysmal SAH dataset, separated by annotated blood volume. Standard deviation between patients reported in parentheses.}\label{tab:unet}
\resizebox{\columnwidth}{!}{%
\begin{tabular}{lrrrrrrr}
\toprule
\multirow{2}{*}
{\makecell[l]{\vspace{2.5ex}Blood Volume}} & (0, 25] & (25, 50] & (50, 100] & (100, 300] & All \\
\cmidrule{2-6}
\makecell[l]{Number of Patients} & 2 & 12 & 7 & 9 & 30 \\
\midrule
None     & 0.107 (0.08) & 0.247 (0.23) & 0.495 (0.17) & 0.629 (0.13) & 0.410 (0.26) \\
Shallow  & \textbf{0.259 (0.10)}  & 0.386 (0.21) & \textbf{0.59 (0.15)}  & 0.681 (0.11) & 0.514 (0.22) \\
Deep     & 0.129 (0.01) & 0.377 (0.22) & 0.554 (0.18) & 0.678 (0.10) & 0.492 (0.23) \\
Encoding & 0.212 (0.12) & 0.359 (0.22) & 0.546 (0.18) & 0.658 (0.11) & 0.483 (0.22) \\
Decoding & 0.200 (0.01)   & \textbf{0.437 (0.19)}  & 0.575 (0.15) & \textbf{0.683 (0.10)}  & \textbf{0.527 (0.20)}  \\
All      & 0.166 (0.01) & 0.399 (0.21) & 0.579 (0.17) & 0.673 (0.12) & 0.508 (0.22) \\
\bottomrule
\end{tabular}
}
\end{table*}

\begin{table*}[t]\centering
\caption{Average patient Dice scores for LoRA and DoRA models applied to Unet on aneurysmal SAH dataset, separated by annotated blood volume. Standard deviation between patients reported in parentheses.}\label{tab:lora}
\resizebox{\columnwidth}{!}{%
\begin{tabular}{lrrrrrrr}
\toprule
\multirow{2}{*}
{\makecell[l]{\vspace{2.5ex}Blood Volume}} & (0, 25] & (25, 50] & (50, 100] & (100, 300] & All \\
\cmidrule{2-6}
\makecell[l]{Number of Patients} & 2 & 12 & 7 & 9 & 30 \\
\midrule
DoRA-C (64)     &    \textbf{ 0.361 (0.09)}&   \textbf{ 0.496 (0.18)}&  0.621 (0.13)    & 0.682 (0.09)&\textbf{0.572 (0.17)}  \\
CP-DoRA (8)        &    0.256 (0.05)&   0.470 (0.20)&   0.620 (0.13) &0.677 (0.10) &0.553 (0.19)  \\
CP-LoRA (64)      &     0.258 (0.02)&   0.480 (0.19) & 0.623 (0.14) &0.683 (0.09)&0.559 (0.19)  \\
LoRA-C (64)       &    0.238 (0.01)&   0.447 (0.21) &  0.618 (0.14) &0.69 (0.10)& 0.546 (0.21) \\
convDoRA (64)   &   0.280 (0.04)&   0.468 (0.19) &   \textbf{0.623 (0.13)} &\textbf{0.694 (0.08)}& 0.559 (0.19)\\
convLoRA (96)    & 0.227 (0.06)&     0.483 (0.18)& 0.617 (0.16)   &0.685 (0.11) & 0.558 (0.19) \\
\bottomrule
\end{tabular}
}
\end{table*}

In fine-tuning Unet-like models, one usually freezes parts of the model, and which parts to freeze has been object of investigation. We describe fine-tuning strategies corresponding to different components of the multi-view Unet in Table \ref{tab:modules} and we present the results of our experiments in Table \ref{tab:unet}, stratified by annotated SAH volume. On the validation set of the pre-training data, the multi-view Unet achieves a mean Dice score of 0.642. We do not report similar results for Unet as they appeared consistently worse than those of the multi-view Unet.

We present the results of our experiments with LoRA/DoRA methods in Table \ref{tab:lora}. For each type of model,  we only present the one with the choice of rank that performed best, among $R=2,4,8,16,32,64,96,128$. The rank is appended to the model name in parentheses (e.g., convDoRA (64) represents the Unet model fine-tuned with the convDoRA method using a rank of 64). As discussed in the methods section, we set the parameter $\alpha=2R$. During fine-tuning, the pretrained Unet weights are completely frozen and only the LoRA/DoRA weights are trained.

\section{Discussion}

After fine-tuning, the multi-view model knowledge translates fairly well to segmenting aneurysmal SAHs at high blood volumes. The best performing strategy i.e. fine-tuning the decoding modules, displays a mean Dice score of 0.683 for volumes $>100mL$. However, it significantly struggles at lower volumes. Notably, at small SAH volumes ($<$25 mL) the Shallow module strategy performs best.

These shallow modules capture low-level features in the CT images \cite{Yao2020-fi}. Because CT images of small aneurysmal SAH are characterized by smaller and more dispersed pockets of blood than TBI cases, we would expect tuning the modules that capture these finer details to perform well. \cite{Amiri2020} suggests that fine-tuning the shallow layers on a Unet model is the best approach for small datasets. However, the small sample size of the SAH cohort limits this analysis. Future work could improve segmentation accuracy for small SAH in several ways including curating a larger dataset from multiple institutions, implementing a more sophisticated hybrid loss function, or implementing post-processing methods.

\begin{figure}
\centering
  
    \includegraphics[width=0.6\columnwidth]{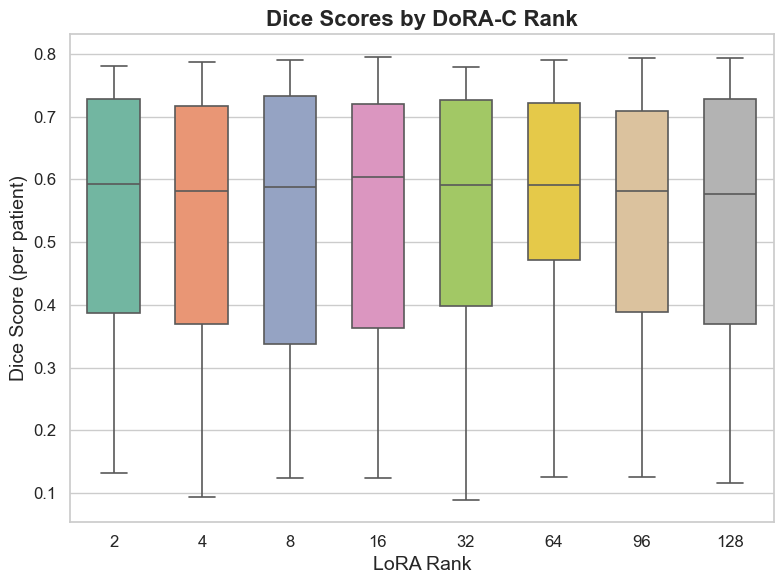}
      \caption{Patient Dice scores of DoRA-C evaluated models on aneurysmal SAH dataset for different choices of rank.}
\label{fig:ranks}
\end{figure}

We see that all the LoRA-based fine-tuned models, applied to a standard Unet, generally outperform standard fine-tuning applied to the multi-view Unet (that outperformed the classic Unet). 
Pairwise comparisons were performed between the LoRA/DoRA models from Table \ref{tab:lora} and the best multi-view model (with decoder fine-tuning) using both a one-sided paired $t$-test and a Wilcoxon signed-rank test (LoRA model $>$ Multiview), as shown in Table \ref{tab:stats}.
Reported Confidence Intervals (CI) correspond to the 95\% CI for the mean paired Dice difference across 30 test patients.
All models except LoRA-C (64) achieved statistically significant improvements after Benjamini--Hochberg correction ($p < 0.05$).

\begin{table}[t]
\centering
\caption{Statistical comparison of LoRA/DoRA models versus the best multi-view model (with decoder fine-tuning). We show 95\% Confidence Intervals (CI) for the paired Dice difference $\Delta$Dice (LoRA/DoRA model $-$ multi-view) and one-sided paired significance tests: $p_t$ (paired \textit{t}-test) and $p_W$ (Wilcoxon signed-rank test), each Benjamini--Hochberg corrected. Significance markers: $^{*}p<0.05$, $^{**}p<0.01$, $^{***}p<0.001$.} \label{tab:stats}
\begin{tabular}{lccc}
\toprule
Model & 95\% CI ($\Delta$Dice) & $p_t$ & $p_W$ \\
\midrule
DoRA-C (64)        & [0.020, 0.076] & $0.0014^{***}$ & $0.0003^{***}$ \\
CP-DoRA (8)          & [0.008, 0.050] & $0.0058^{**}$  & $0.0019^{**}$  \\
CP-LoRA (64)         & [0.016, 0.056] & $0.0014^{***}$ & $0.0003^{***}$ \\
LoRA-C (64)        & [--0.003, 0.046] & $0.0400^{*}$  & $0.0249^{*}$  \\
convDoRA (64)   & [0.014, 0.057] & $0.0014^{***}$ & $0.0010^{**}$  \\
convLoRA (96)   & [0.016, 0.051] & $0.0014^{***}$ & $0.0003^{***}$ \\
\bottomrule
\end{tabular}
\end{table}

Significant gains in performance are obtained at volumes $<100mL$.
The best overall LoRA-based fine-tuned model DoRA-C (64) outperforms the best multi-view fine-tuned model at volumes $<100mL$, while maintaining the same performance at volumes $>100mL$.
At high volumes, the best LoRA-based fine-tuned model convDoRA (64) outperforms the best multi-view fine-tuned model at 0.694 (0.08) compared to 0.683 (0.1).

\begin{figure}[t]
\begin{center}
    \includegraphics[width=\columnwidth]{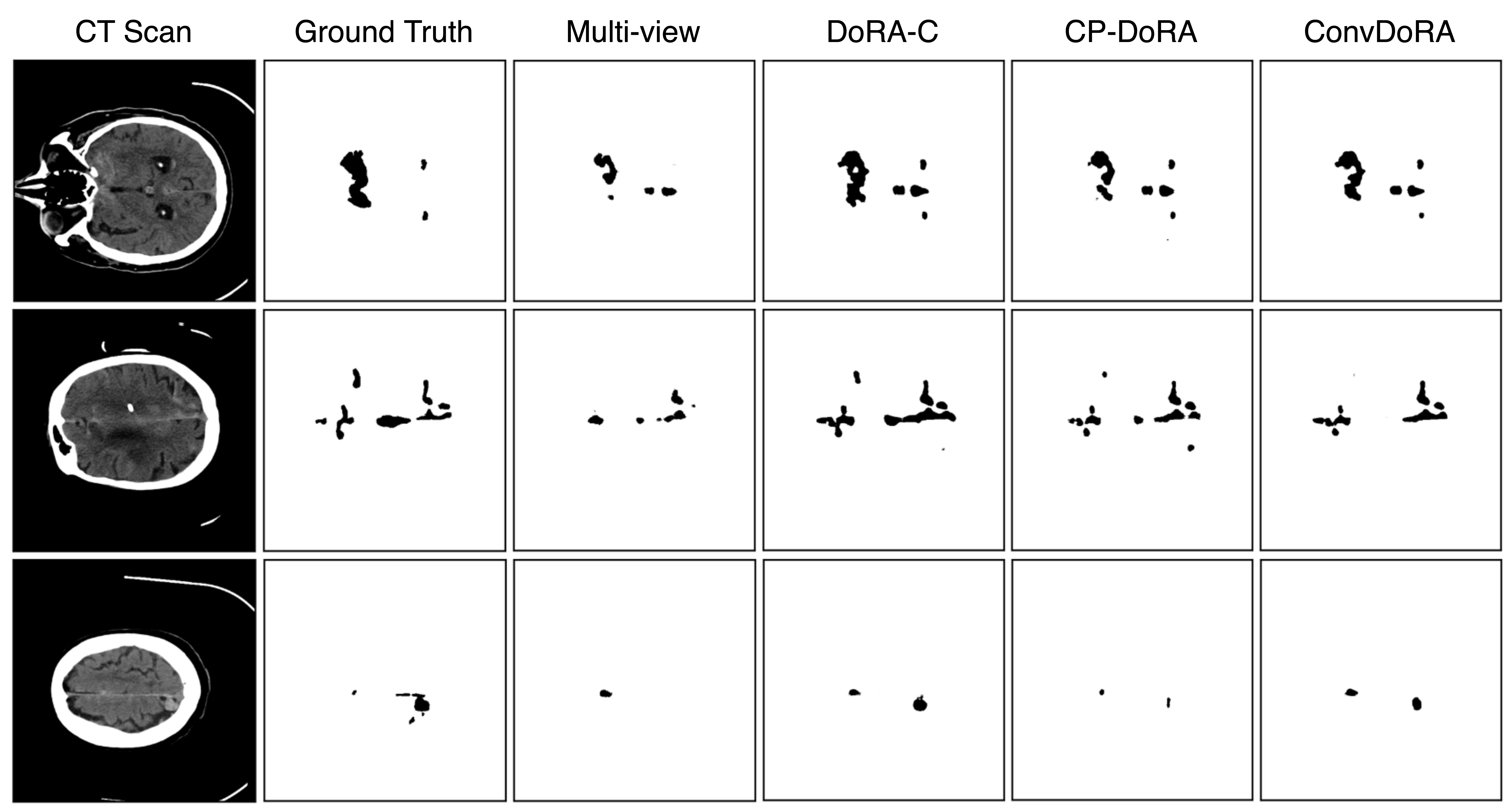}
    \caption{Segmentation masks produced by different models. LoRA/DoRA methods use rank 64.}
\end{center}
\label{fig:masks}
\end{figure}

We show in Figure \ref{fig:ranks} how the best model's performance varies with the rank. While one can achieve good performance at lower ranks for a fraction of the parameters, our analysis shows that using larger ranks and over-parametrizing the model leads to both best average performance and smaller variance. This behavior is analogous to another deep learning phenomenon known as double descent \cite{doubledescent}. With the exception of DoRA, all models perform best at rank 64, or even 96 in the case of convLoRA. The model with the best overall performance is DoRA-C at rank 64, due to a significantly better segmentation at lower volumes compared to other models; see Figure \ref{fig:masks} for example segmentations of this and other models. However, at high volumes specifically DoRA-C is outperformed by convDoRA. Notably, all these LoRA/DoRA models perform better than standard Unet fine-tuning (no matter the fine-tuning strategy) both in terms of higher average and reduced standard deviation. We hypothesize that the reduced standard deviation results from LoRA/DoRA approaches regularizing model complexity in accordance with the bias-variance tradeoff for neural networks \cite{Neal2018-ah}.

Model performance varies across SAHs of different volume. All fine-tuning strategies showed increasing performance as the volume of the SAH increases. This is likely due to the more severe cases have more distinct visual features as well as small errors have a greater impact on the Dice score of small SAHs. Hemorrhage volume is an important indicator of later complications, with a larger volume increasing severity and risk. At higher volumes ($>100mL$), convDoRA at rank 64 perform best with a mean Dice score of 0.694 (and smallest standard deviation).

\begin{figure}[t]
\begin{center}
    \includegraphics[width=0.6\columnwidth]{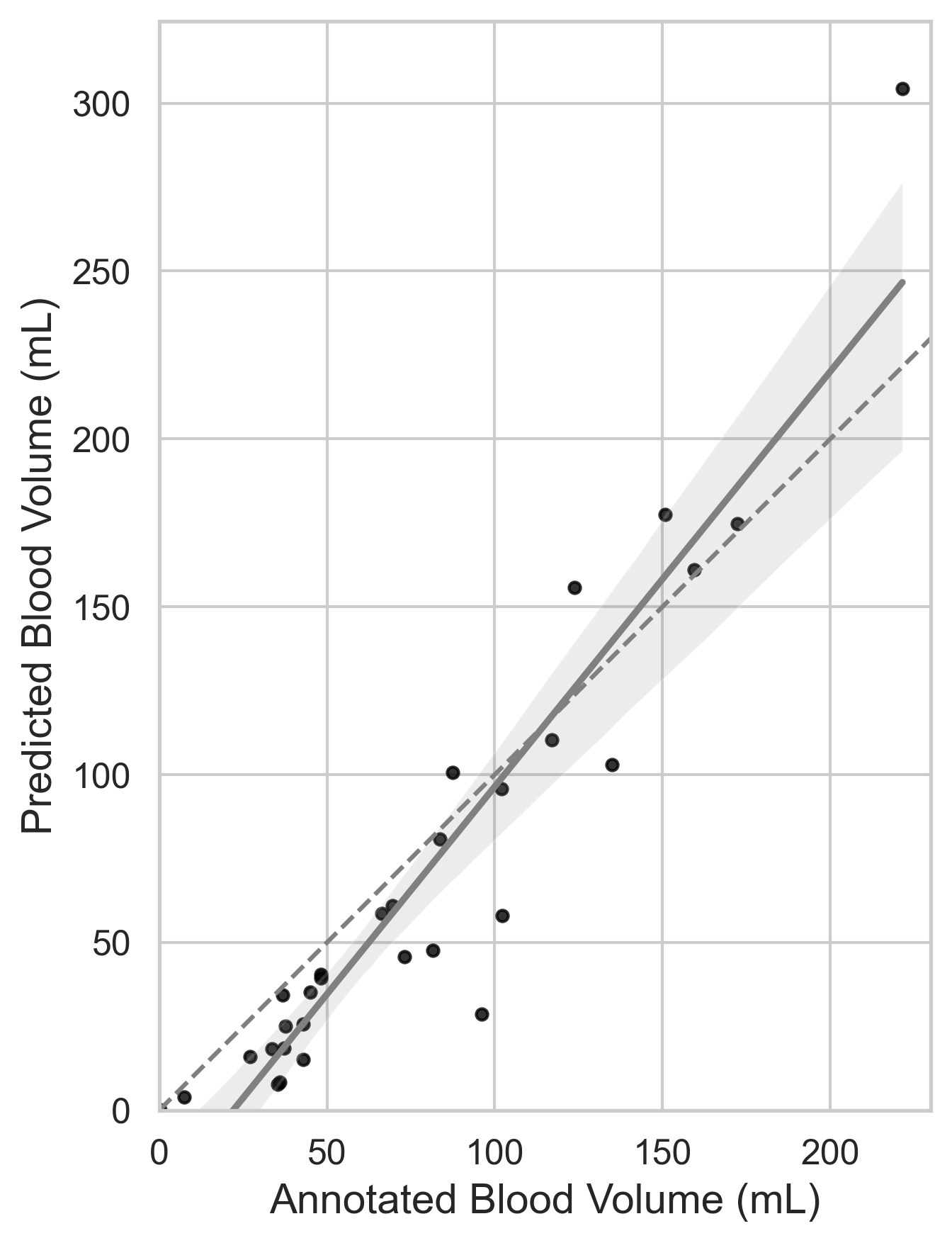}
    \caption{Predicted and annotated SAH volumes from the DoRA-C model with rank 64. The solid line is the fitted linear regression with 95\% confidence intervals and the dashed one is the calibration line.}
    \label{fig:volume}
    \end{center}
\end{figure}

As shown in Figure \ref{fig:volume}, DoRA-C at rank 64 tended to underestimate blood volume in moderate cases ($50$-$100mL$), and overestimate in severe cases ($>100mL$). These discrepancies may be a result of differences between the pre-training TBI and tuning SAH sets. With further model refinement, these estimated SAH volumes may aid clinicians in assessing patient risk for clinical outcomes like vasospasm \cite{Sanchez2025-oi}.

Overall, this work shows the potential of leveraging large, multi-institutional TBI datasets to solve clinical tasks on less common and more specific pathologies like aneurysmal SAH.

While CP-LoRA and CP-DoRA do not achieve best performance, they often score similarly while employing significantly fewer parameters (30-40\%): for each layer, the number of parameters used in fine-tuning is ${(c_{in}+c_{out}+2k)R}$. For a complete overview of the number of trainable parameters see Table \ref{tab:params}. Therefore these methods might still prove valuable in settings with limited resources, in particular for larger Unet models. For example, \cite{Kiewitz2024-mk} reports worse performance from 3D Unet than 2D at aneurysmal SAH segmentation, possibly due to insufficient training samples for the 3D model with significantly more parameters. In such cases, CP-LoRA and CP-DoRA could be especially valuable and easily extended.

These results suggest automated SAH segmentation via transfer learning as a plausible alternative to time-consuming manual annotation. On a different aneurysmal SAH data, manual annotators achieved an average Dice score of 0.64 \cite{Boers2014-mi}. However, that dataset contained, on average, smaller SAHs: 39.71 $\pm$ 32.84 mL compared to 77.43 $\pm$ 52.24 mL in our dataset. On severe aneurysmal SAH cases, the proposed fine-tuning strategy may perform on par with clinicians at a fraction of the time. We note that this study has several limitations including using a small SAH dataset from a single institution which limits generalizability to other healthcare centers, demographics, and heterogeneous severity and disease presentations. Further development and validation would be required prior to algorithmic deployment in a clinical setting.

\section{Conclusions}

This paper makes several contributions. First, it evaluates the feasibility of transfer learning across different hematoma types, namely TBI-induced hematomas to SAH,  using limited training data. Second, it introduces a novel low-rank adaptation method for Unet, called CP-LoRA, which achieves parameter efficiency compared to existing approaches such as LoRA-C and convLoRA. Third, it proposes DoRA-based variants of LoRA methods, namely DoRA-C, convDoRA, and CP-DoRA, incorporating orthogonal regularization into low-rank adaptation. Finally, the paper presents a comprehensive comparison of all LoRA and DoRA variants in the context of a Unet architecture, showing that overparameterization can lead to improved segmentation performance relative to strictly low-rank adaptations. Overall, the results show that LoRA-based methods outperform standard Unet fine-tuning strategies even on a more powerful multi-view architecture, and that transfer learning with limited data between hematoma types is a promising application.

\backmatter

\backmatter

\section*{List of abbreviations}

\noindent CNN - Convolutional Neural Network

\noindent CP - Canonical Polyadic

\noindent DoRA - Weight-Decomposed Low-Rank Adaptation

\noindent LoRA - Low Rank Adaptation

\noindent SAH - Subarachnoid Hematoma

\noindent  TBI - Traumatic Brain Injury

\section*{Declarations}

\begin{itemize}
\item Conflict of interest/Competing interests: CW is an Unaffiliated Neurotrauma Consultant with the National Football League. 
\item Ethics approval and consent to participate: The studies involving de-identified human subject data was reviewed and approved by University of Michigan Institutional Review Board (HUM00160386).  Written informed consent from patients has been waived by the University of Michigan Institutional Review Board since this study involves no more than minimal risk to the involved subjects. All methods were performed in accordance with the relevant guidelines and regulations. 
\item Consent for publication: Not applicable.
\item Funding: No funding to declare.
\item Availability of data and materials: Part of the data that support the findings of this study are available from Progesterone for Traumatic Brain Injury Experimental Clinical Treatment (ProTECT) III Trial, but restrictions apply to the availability of these data. All other datasets from the current study were collected at Michigan Medicine. The University of Michigan’s Innovation Partnerships (UMIP) unit will handle potential charges/arrangements of the use of data by external entities, using such methods as material transfer agreements. Please contact UMIP (innovationpartnerships@umich.edu) for data inquiries.
\item  The code used for this study is available at https://github.com/Minoch/LoRA-based-methods-SAH-segmentation.
\item Authors' contributions: data acquisition - RM, CW, EW; data annotation - RM, CW; project ideation - CM, KN, CW; methodology - CM; project supervision - CM, KN, EW; data preprocessing: HM, MH, CM; software: CM, MH, HM; formal analysis and validation: CM, MH; first draft: CM, MH; final revision: CM, MH, KN, EW.
\end{itemize}


\bibliographystyle{bst/sn-nature}
\bibliography{sn-bibliography}

\end{document}